\documentclass[prd,showkeys,floatfix,nofootinbib,
               preprint,12pt,
               tightenlines,fleqn]{revtex4} %for preprint

\usepackage{amsmath,amssymb,revsymb,graphicx,dcolumn}
\usepackage{leftidx}
\usepackage{slashed}

\usepackage{slashed}

\newcommand{\beq}{\begin{equation}}
\newcommand{\eeq}{\end{equation}}
\newcommand{\beqa}{\begin{eqnarray}}
\newcommand{\eeqa}{\end{eqnarray}}
\newcommand{\bsubeqs}{\begin{subequations}}
\newcommand{\esubeqs}{\end{subequations}}

\usepackage{mwe}

\begin{document}
\hfill KA--TP--06--2017\vspace*{4mm}\newline
\title[]
      {Can a QED photon propagate out of the black-hole horizon?}
\author{Slava Emelyanov}
\email{viacheslav.emelyanov@kit.edu}
\affiliation{Institute for Theoretical Physics,\\
Karlsruhe Institute of Technology (KIT),\\
76131 Karlsruhe, Germany\\}

\begin{abstract}
\vspace*{2.5mm}\noindent
We show at one-loop approximation that the QED photon acquires a negative gauge-invariant
mass-squared in the near-horizon region of an evaporating Schwarzschild black hole. This result may
imply that information about the inner structure of evaporating black holes is accessible, even without
invoking the modification of the basic QFT and GR principles.
\end{abstract}


\maketitle

\section{Introduction}

Physical black holes are expected to evaporate in quantum theory~\cite{Hawking}. The evaporation
process is, however, featureless. This circumstance leads to the information loss in black-hole physics.
The outstanding question is if a semi-classical mechanism exists which allows to access the information
fell in black holes (for a recent critical review of various proposals towards the resolution of the
information-loss problem, see~\cite{Unruh&Wald}).

Inspired by the proposal of non-violent nonlocality near the event horizon of black
holes~\cite{Giddings-1,Giddings-2}, we intend to study in this article the photon dispersion
relation at one-loop order in quantum electrodynamics (QED) near the horizon, with the
black-hole-evaporation effect taken into account.

Through this article, we take $c=G=k_B = \hbar = 1$, unless otherwise stated.

\section{Photon mass-squared}

In the Minkowski vacuum in the absence of external fields, the photon dispersion relation reads
$\omega_\mathbf{k} = |\mathbf{k}|$ at any order of perturbation theory due to the
gauge and Lorentz symmetries, where $\omega_\mathbf{k}$ and $\mathbf{k}$ are,
respectively, the photon's energy and momentum. This relation typically changes in
mediums/non-trivial quantum states and/or in the presence of external fields. In particular, the
photon gains a gauge-invariant mass-squared $m_\gamma^2$ equaling to $\frac{1}{6}e^2T^2$
in the neutral electron-positron plasma of temperature $T \gg m_e$, where $e > 0$ is the
elementary charge and $m_e$ is the electron mass. Specifically, it turns out that
$\omega_\mathbf{k} \approx (|\mathbf{k}|^2 + m_\gamma^2)^\frac{1}{2}$ if the
condition $eT \ll |\mathbf{k}| \ll T$ is fulfilled~\cite{Weldon}. In case of the cold plasma, 
$T \ll m_e$, the thermal photon mass is still non-vanishing, but suppressed by the
Boltzmann factor $\exp(-m_e/2T)$.

The method used in~\cite{Weldon} to compute $m_\gamma$ is diagrammatic, i.e. based on
the computation of the photon self-energy at one-loop order in perturbation theory. The same
result can be derived within the kinetic-theory approach~\cite{Blaizot&Iancu}, as these
turn out to be equivalent~\cite{Mrowczynski&Thoma}. We shall employ below both methods.

The local quantum physics in the near-horizon region of an evaporating Schwarzschild
black hole is described by the following Feynman propagator of the massless Dirac field:
\begin{equation}
S(x,x') = S_M(x,x') + S_R(x,x')\,
\end{equation}
with $S_M(x,x')$ locally corresponding to the Minkowski propagator and $S_R(x,x')$ being a
non-singular solution of the Dirac equation, which gives the evaporation effect:
\begin{equation}\label{eq:sr}
S_R(x,x') \approx -\frac{27}{32\pi^2}{\int\limits_0^{+\infty}}\frac{dp\,p}{e^{\beta p} + 1}\,
i\slashed{\nabla}\cos\big(p\Delta{v}\big).
\end{equation}
where $\beta \equiv 4\pi r_H \equiv 1/T_H$ is the inverse Hawking temperature~\cite{Hawking}
with the black-hole Schwarzschild radius $r_H$, and $v \equiv t + r_*$ is the in-going
Eddington-Finkelstein coordinate, where the Regge-Wheeler radial coordinate $r_*$ is equal
to $r + r_H\ln(r/r_H - 1)$. It should be pointed out that  we have employed in~(\ref{eq:sr}) the
DeWitt approximation for the transmission amplitude, which explains the factor $27$~\cite{DeWitt}.
Moreover, the basic structure of~(\ref{eq:sr}) can be envisaged from the fact that
$g^{\mu\nu}\nabla_\mu\nabla_\nu \propto (\partial/\partial{u})(\partial/\partial{v})$ holds close
to the horizon, where $u \equiv t  - r_*$ is the out-going Eddington-Finkelstein coordinate, as
well as the non-singularity condition for $S_R(x,x')$ in the coincidence limit $x' \rightarrow x$,
and that~(\ref{eq:sr}) has to give non-trivial stress-energy tensor, which does not diverge on
the horizon.

In the Fermi normal coordinates $\bar{x} \equiv (\tau,\mathbf{x}) \equiv (\tau,x,y,z)$,
we have $\Delta{v} \approx \frac{1}{2}(\Delta\tau + \Delta{x})$, where the local region
$x > \tau\,(x < \tau)$ is outside (inside) the black-hole horizon. Replacing $p$ by $2p$
in~(\ref{eq:sr}) and, then, rewriting~(\ref{eq:sr}) in terms of the Fermi normal coordinates,
we obtain the following result:
\begin{equation}\label{eq:sr-lmf}
S_R(\bar{x},\bar{x}') \approx -\frac{27}{8\pi^2}{\int\limits_0^{+\infty}}\frac{dp\,p}{e^{2\beta p} + 1}\,
i\slashed{\partial}\cos\big(p(\Delta\tau + \Delta{x})\big)\,,
\end{equation}
which holds in the local Minkowski frame freely falling near the black-hole horizon. The form
of $S_R(\bar{x},\bar{x}')$ is valid for any value of $p$, assuming that $|\bar{x}-\bar{x}'| \ll r_H$ holds.

\subsection{Quantum-statistical approach}

In the quantum-kinetic-theory approach~\cite{Emelyanov-17-a,Emelyanov-17-b},
$S_R(\bar{x},\bar{x}')$ in~(\ref{eq:sr-lmf}) gives the following ``one-particle" distribution:
\begin{eqnarray}\nonumber\label{eq:opd}
f_{e}(\mathbf{x},\mathbf{p}) \hspace{-1mm}&=&\hspace{-1mm} f_{\bar{e}}(\mathbf{x},\mathbf{p})
\\[1mm]
\hspace{-1mm}&\approx&\hspace{-1mm} -\frac{27}{8\pi^2}\frac{|\mathbf{p}|^2}{e^{2\beta|\mathbf{p}|} +1}
\,\theta(-p_x)\delta(p_y)\delta(p_z)
\end{eqnarray}
with the on-mass-shell condition $p_0 = |\mathbf{p}|$.~This distribution function
$f_{e/\bar{e}}(\mathbf{x},\mathbf{p})$ refers to the massless electron/positron field,
where the momentum space associated with the point $(\tau,\mathbf{x})$ is defined
according to~\cite{Bunch&Parker}.

Physically, the distribution function $f_{e/\bar{e}}(\mathbf{x},\mathbf{p})$
describes an inward negative-energy-density flux (due to the electron/positron
field) into the event horizon, which leads to $\dot{r}_H < 0$, where the dot
stands for the differentiation with respect to the Schwarzschild time. This can be directly
shown by calculating the stress-energy tensor in the local Minkowski/Fermi frame:
\begin{equation}\label{eq:set}
\langle T_{e\bar{e}}^{\mu\nu}(\bar{x}) \rangle =
{\int}\frac{d^3\mathbf{p}}{|\mathbf{p}|}\,p^\mu p^\nu
\big(f_{e}(\mathbf{x},\mathbf{p}) + f_{\bar{e}}(\mathbf{x},\mathbf{p})\big)\,.
\end{equation}

The negative sign of $f_{e/\bar{e}}(\mathbf{x},\mathbf{p})$ is non-typical for classical many-particle
systems, but consistent with the computation of the stress-energy tensor near the horizon in the
Schwarzschild frame~\cite{Candelas}. This is also in agreement with the fact that the inward
quantum flux of the negative energy density originates well outside of the event horizon~\cite{Unruh}.
This means that this flux cannot be understood in classical terms~\cite{Emelyanov-17-a}.

Employing the kinetic-theory approach for the computation of one-loop effects in quantum
electrodynamics~\cite{Blaizot&Iancu}, we obtain in the near-horizon region that
\begin{equation}\label{eq:pms}
m_\gamma^2 = 2e^2{\int}\frac{d^3\mathbf{p}}{|\mathbf{p}|}
\big(f_{e}(\mathbf{x},\mathbf{p}) + f_{\bar{e}}(\mathbf{x},\mathbf{p})\big)
\approx -\frac{27}{16}\frac{e^2T_H^2}{6}\,,
\end{equation}
where we have used $f_{e/\bar{e}}(\mathbf{x},\mathbf{p})$ given in~(\ref{eq:opd}).
The negative value of $m_\gamma^2$ is the main result of this article.

\subsection{Diagrammatic approach}

The photon mass-squared $m_\gamma^2$ can also be obtained by computing
the one-loop photon self-energy, which provides the lowest-order radiative correction to the photon
propagator. To this purpose, we need the electron-positron propagator in the
momentum-space representation near the black-hole horizon:
\begin{equation}\label{eq:fp}
S(\bar{x},\bar{x}') = {\int}\frac{d^4p}{(2\pi)^4}\,\Big(S_M(p) + S_R(p)\Big)
e^{-ip(\bar{x}-\bar{x}')}\,,
\end{equation}
with the definitions
\begin{eqnarray}
S_M(p) \hspace{-1mm}&\approx&\hspace{-1mm} \frac{i\slashed{p}}{p^2 + i\varepsilon}\,,
\\[1mm]
S_R(p) \hspace{-1mm}&\approx&\hspace{-1mm} 27\pi^2\frac{|\mathbf{p}|\,\slashed{p}}{e^{2\beta|p_0|} + 1}\,
\delta\big(\mathbf{p} + p_0\mathbf{n}\big)\,,
\end{eqnarray}
where $\mathbf{n} \equiv (1,0,0)$ is an outward unit three-vector being locally identical
with the radial direction from the centre of the black hole~\cite{Emelyanov-17-a,Emelyanov-17-b}.

With the identity
\begin{equation}
\langle \textrm{tr}\big(\bar{\psi}\gamma_\mu\partial_\nu\psi\big) \rangle = 
-\lim_{\bar{x} \rightarrow \bar{x}'}
\textrm{tr}\big(\gamma_\mu\partial_\nu S(\bar{x},\bar{x}')\big)\,, 
\end{equation}
one can derive a part of the renormalised stress-energy tensor, which originates from the second
term in round
brackets on the right-hand side of~(\ref{eq:fp}). This gives rise to the inward flux of the
negative energy density, in complete agreement with~(\ref{eq:set}). The rest part
of the total stress tensor comes from the trace anomaly,
which arises from the renormalization, that could be based, e.g., on the Hadamard-parametrix
subtraction.

Furthermore, the correction to the Minkowski propagator in~(\ref{eq:fp}) leads to the
change of the pole structure of the photon propagator. It should be noted that
$S_M(p)$ alone does not change this structure at any order of perturbation theory due
to its Lorentz-invariant form and the gauge symmetry. Following~\cite{Weldon}
and~\cite{Emelyanov-17-b}, we obtain at one-loop approximation that
\begin{equation}\label{eq:pms-a}
m_\gamma^2 \approx \lim_{|\mathbf{k}| \rightarrow \omega_\mathbf{k}}{\int\limits_0^{+\infty}}
\frac{dp\,p^3}{e^{2\beta p}{+}1}\,\Pi(k,p)  = -\frac{27}{16}\frac{e^2T_H^2}{6}\,,
\end{equation}
where
\begin{equation}
\Pi(k,p) \equiv -\frac{27e^2}{\pi^2}
\frac{4\omega_\mathbf{k}|\mathbf{k}|\cos\theta+(\omega_\mathbf{k}^2+\mathbf{k}^2)(1+\cos^2\theta)}
{4p^2(\omega_\mathbf{k}+|\mathbf{k}|\cos\theta)^2 - (\omega_\mathbf{k}^2-\mathbf{k}^2)^2}\,,
\end{equation}
where $\theta$ is the angle between $\mathbf{n}$ and $\mathbf{k}$. The equation~(\ref{eq:pms-a})
is, thus, identical to~(\ref{eq:pms}), in accordance with~\cite{Mrowczynski&Thoma}. This
implies that the result $m_\gamma^2 \sim -e^2T_H^2$ is accurate within the approximation considered.

For the one-loop result~(\ref{eq:pms}) to be reliable, the de Broglie wavelength
of the photon must be much less than the black-hole size $r_H$, otherwise no notion of the
photon exists in the Wigner sense, which appears to be the only one that is \emph{a posteriori}
physically relevant in elementary particle physics. In other words, the photon, we consider herein,
is high-energy enough (with energy being still very much less than the Planck energy
$(\hbar c^5/G)^\frac{1}{2} \approx 1.2{\times}10^{19}\,\textrm{GeV}$). The energy scale of order
$1/r_H$ provides then a natural IR cutoff below which the one-loop computation is
unreliable.\footnote{The IR cutoff is necessarily present, if the Minkowski-space approximation
is used, but locally immaterial in particle physics, because particles in QFT are localized field
excitations.}

In fact, the electron-positron field is not massless in nature. Nevertheless, the
result~(\ref{eq:pms}) holds for black holes of mass $M \ll 10^{16}\,\textrm{g}$,
because then $m_e \ll T_H$ is fulfilled, i.e. the electron-positron field can be
considered as effectively massless (the hard-thermal-loop approximation).
We obtain that $m_\gamma^2$ is suppressed by the Boltzmann factor for larger
black holes, but $m_\gamma^2$ is still non-vanishing and negative.

\section{Physical consequence of $m_\gamma^2 < 0$}

In order to find out the physical consequence of the modified dispersion relation,
$\omega_\mathbf{k} = (\mathbf{k}^2 - |m_\gamma^2|)^\frac{1}{2} > 0 $, one needs to consider
a wave package $\psi_\gamma$, which describes the real photon with the momentum
$|\mathbf{k}| \gg 1/r_H$ in the space-time region with $\textrm{supp}(\psi_\gamma) \neq \varnothing$
close to the black-hole horizon.

In quantum theory, this wave package gives rise to the following one-photon state:
\begin{equation}
|\psi_\gamma(\bar{x})\rangle = {\int}\frac{d^3\mathbf{p}}{(2\pi)^3}\frac{1}{2\omega_\mathbf{p}}\,
e^{ip\bar{x}}\gamma_\mathbf{k}(\mathbf{p}) |\mathbf{p}\rangle\,,
\end{equation}
where the photon-polarisation index has been suppressed. The state
$|\mathbf{p}\rangle \equiv \sqrt{2\omega_\mathbf{p}}\hat{a}_\mathbf{p}^\dagger|vac\rangle$ is the
momentum-operator (non-normalisable) eigenstate and
\begin{equation}
\gamma_\mathbf{k}(\mathbf{p}) =
N\exp\left[{-}\frac{(\mathbf{p} - \mathbf{k})^2}{4\sigma}\right]
\end{equation}
with the normalisation factor $N$ and the momentum variance $\sigma$. In what
follows, we choose that $\sigma \ll |\mathbf{k}|^2$ in order to have the photon of
the momentum $\mathbf{k}$. The momentum variance $\sigma$ implies that
$|\psi_\gamma(\bar{x})\rangle$ is described by the position variance of order $1/\sigma$ in the vicinity
of the point $\bar{x}$, in
accordance with the Heisenberg uncertainty relation. Assuming that $r_H|\mathbf{k}| \gg 1$,
we obtain $|\psi_\gamma(\bar{x})\rangle$ is localised at $\bar{x}$ with the support
of size much less than $r_H$. This all means that $|\psi_\gamma\rangle$ (with the unspecified
position in spacetime) is the one-photon
state which is sufficiently localised near the black-hole horizon of radius $r_H$, such that we
are allowed to apply the powerful machinery of QFT used in elementary particle physics to
the photon $|\psi_\gamma\rangle$.

The position-space probability amplitude for the photon to propagate from $\bar{x}'$ to $\bar{x}$ reads
\begin{equation}\label{eq:position-space-amplitude}
A_\gamma(\bar{x},\bar{x}') = \langle \psi_\gamma(\bar{x})|\psi_\gamma(\bar{x}')\rangle\,,
\end{equation}
which reaches its maximal value if the photon propagates along the classical trajectory:
$\mathbf{x} = \mathbf{x}' + (\mathbf{k}/\omega_\mathbf{k})(\tau - \tau')$. The space-time
interval $s(\bar{x},\bar{x}')$ is then given by
\begin{equation}\label{eq:space-time-interval}
\eta_{\mu\nu}(\bar{x}-\bar{x}')^\mu (\bar{x}-\bar{x}')^\nu \approx
-\frac{|m_\gamma^2|}{\mathbf{k}^2}(\tau - \tau')^2\,,
\end{equation}
where we have taken into account that $|\mathbf{k}| \gg |m_\gamma|$ for the photon
described by $|\psi_\gamma\rangle$. Since
$(\bar{x} - \bar{x}')^2 < 0$ at one-loop level near the event horizon, it is tempting to
conclude that this high-energy photon may be able to propagate out of the black-hole horizon
if it was initially inside.

The result~(\ref{eq:space-time-interval}) is in apparent tension with the causality
in QED. The causality principle belongs to one of the postulates in the Wightman axiomatic approach to
QFT~\cite{Haag}. Specifically, the notion of the Einstein causality in local quantum field theory refers
to the circumstance that two local
observables $\hat{O}(\bar{x})$ and $\hat{O}(\bar{x}')$ commute
with each other if these are space-like separated, i.e. $(\bar{x}-\bar{x}')^2 < 0$.
But the commutator $[\hat{O}(\bar{x}),\hat{O}(\bar{x}')]$ is generically not
a c-number in interacting QFTs~\cite{Haag} and, thus, depends on a quantum state under consideration.

As an example, one can choose $\hat{O}(\bar{x})$ be the particle-number
operator $\hat{N}(\bar{x})$, which gives the number of photons in the vicinity of the
point $\bar{x}$. This means that
\begin{equation}
\langle \psi_\gamma|\hat{N}|\psi_\gamma\rangle =
\left\{
\begin{array}{ll}
1\,, &\hspace{0mm}\textrm{supp}(N)\cap\textrm{supp}(\psi_\gamma) \neq \varnothing\,, \\[3mm]
0\,, &\hspace{0mm}\textrm{supp}(N)\cap\textrm{supp}(\psi_\gamma) = \varnothing\,.
\end{array}
\right.
\end{equation}
In practice, $\hat{N}$ corresponds to a certain single-photon detector (SPD), which clicks if the
photon and the detector have overlapping supports.\footnote{In case of charged particles,
$\hat{N}$ can be realised, for instance, with the help of the Wilson cloud chamber, where
$\textrm{supp}(N)$ corresponds to the space-time domain where that is located. A charged
particle can then be observed if it penetrates inside the chamber. On the contrary, if this
charged particle does not pass through the space-time region where the cloud chamber is,
then the particle remains unobserved.} In the vacuum state $|vac\rangle$, one has that
\begin{equation}
\langle vac|[\hat{N}(\bar{x}),\hat{N}(\bar{x}')]|vac\rangle = 0\,.
\end{equation}
The result is trivial for any $\bar{x}$ and $\bar{x}'$, because there are no real photons in
$|vac\rangle$ and SPD cannot, therefore, click. On the other hand, one can consider the
following state:
\begin{equation}
|hor\rangle =
\left\{
\begin{array}{ll}
|vac\rangle\,, & \bar{x} \,\notin\, \textrm{supp}(\psi_\gamma)\,, \\[2mm]
|\psi_\gamma\rangle\,, & \bar{x} \,\in\, \textrm{supp}(\psi_\gamma)\,.
\end{array}
\right.
\end{equation}
This describes a quantum state in the near-horizon region, corresponding to one photon of the
momentum $\mathbf{k}$, which hovers somewhere near
the black-hole horizon. In the non-vacuum state $|hor\rangle$, one has that
\begin{equation}
\langle hor|[\hat{N}(\bar{x}),\hat{N}(\bar{x}')]|hor\rangle = 
A_\gamma(\bar{x},\bar{x}')\,,
\end{equation}
where we have assumed that both SPDs located at $\bar{x}$ and $\bar{x}'$ click.
Normally, $A_\gamma(\bar{x},\bar{x}')$ is essentially zero if $\bar{x}$ and $\bar{x}'$
are space-like separated, which holds for any particle from the Standard Model.
That is what we experience on earth. In the near-horizon region,
$A_\gamma(\bar{x},\bar{x}')$ gets, however, its maximum for $\bar{x}$ and $\bar{x}'$, which are slightly
space-like separated as follows from~(\ref{eq:position-space-amplitude})
and~(\ref{eq:space-time-interval}).\footnote{In the electron-positron plasma, $A_\gamma(\bar{x},\bar{x}')$
is, in contrast, essentially non-vanishing for time-like separated $\bar{x}$ and $\bar{x}'$, because
the effective QED-photon mass is real there~\cite{Weldon}.}
This means that the Einstein causality may be violated
near the horizon of evaporating Schwarzschild black holes, as a result of the non-classical
quantum properties of $|vac\rangle$ at $r \sim r_H$.

On the other hand, it is usually expected that the negative (Lagrangian) mass-squared $-\mu^2 < 0$
inevitably leads to the vacuum instability. The prime example is the spontaneous symmetry 
breaking in the standard electroweak theory. This happens, because the Higgs-field modes with
$|\mathbf{k}|^2$ less than the negative-mass-squared absolute value, $\mu^2$,
correspond to imaginary-frequency modes, which grow exponentially with
time. But if these imaginary-frequency modes are not taken in the quantum-field expansion,
then the quantum vacuum is stable, but the theory is non-causal~\cite{Aharonov&Komar&Susskind}.

In the case under consideration, we have that $\mu = |m_\gamma|$, where
$|m_\gamma| \sim eT_H$ holds, i.e. the origin of $\mu \neq 0$ is quantum in nature as, first,
$T_H \propto \hbar$ and, second, stems from one-loop order in perturbation theory, which
is non-existent in the classical limit, i.e. $\hbar \rightarrow 0$.
Moreover, $1/\mu$ is much larger than the curvature length near the horizon, which is of the order of $r_H$.
For two space-time points $\bar{x}$ and $\bar{x}'$ with $|\bar{x} - \bar{x}'| \sim r_H$,
the local curvature is non-negligible and, thus, the local Minkowski geometry is no longer a
good approximation, e.g.
\begin{equation}
\eta_{\mu\nu} \sim R_{\mu\lambda\nu\rho}(\bar{x}')(\bar{x}-\bar{x}')^\lambda(\bar{x}-\bar{x}')^\rho
\end{equation}
for $|\bar{x} - \bar{x}'| \sim r_H$ near the horizon. This means that the notion of the photon
described by the wave package of size of order $r_H$ and larger is meaningless within the framework of
elementary particle physics in the near-horizon region. Since the one-loop result
$m_\gamma^2 \neq 0$ has been obtained within that framework, it cannot be applied to those low-energy
photons. These photons cannot even be defined as belonging to one of the unitary and irreducible
representations of the Poincar\'{e} group, which is a isometry group of the local Minkowski geometry.
In general, the Lehmann-Symanzik-Zimmermann reduction formula relating physical particle states
with poles in the Feynman propagators is then no more working and, thus, elementary particle physics
becomes physically obscure.

Consequently, the perturbative result $m_\gamma^2 < 0$ does not
imply that the quantum-vacuum instability, that normally happens due to the frequency
modes with $|\mathbf{k}| < \mu$ in theories with the negative (Lagrangian) mass-squared term,
has to occur in the case considered,
because the one-loop computation providing $m_\gamma^2 \neq 0$
is not applicable to wave packages with $r_H|\mathbf{k}| < 1$. But $m_\gamma^2 \neq 0$
is applicable to the high-energy photons, which are described by wave
packages of the type $|\psi_\gamma\rangle$.\footnote{It is worth noting that this observation also agrees
with the expectation that radiative corrections should be small in perturbation theory. In particular, this means
that the dispersion relation in perturbation theory can be only marginally modified: $\omega_\mathbf{k} =
(|\mathbf{k}|^2 - |m_\gamma^2|)^\frac{1}{2} = |\mathbf{k}| + \textrm{O}(e^2)$.} This circumstance also
appears to exclude an extremely rapid black-hole-instability scenario, which occurs in field theories
with the faster-than-light phase velocity~\cite{Barbado&Barcelo&Garay&Jannes}
(incidentally, the result~(\ref{eq:pms}) leads to the slower-than-light phase velocity).

If the causality-violation effect happens in nature near evaporating black holes, then the
real photons we could be interested in must not be ultra-high-energy ones in
order to be able to propagate out of the black-hole horizon. This extra condition appears from the expression
for the photon group velocity, $\mathbf{k}/\omega_\mathbf{k}$, for which we have that
\begin{equation}
\frac{|\mathbf{k}|}{\omega_\mathbf{k}} - 1 \approx
\left\{
\begin{array}{ll}
10^{-6}\,, & r_H|\mathbf{k}| \rightarrow 10\,, \\[2mm]
0\,, & r_H|\mathbf{k}| \rightarrow \infty
\end{array}
\right.
\end{equation}
in the case of primordial evaporating
black holes of today's mass $M \ll 10^{16}\,\textrm{g}$. For astrophysical black holes, the upper
bound for the photon group velocity is extremely close to unity. Therefore, the Einstein acausality,
in favour of which our result turns out to decide, should not imply any tension with
the up-to-date observations on earth.

\section{Conclusions}

As a consequence of the black-hole evaporation, the high-energy-enough QED photon acquires a
gauge-invariant mass-squared at one-loop order in perturbation theory. The mass-squared
$m_\gamma^2$ turns out to be of order $e^2T_H^2$ and negative. The latter is directly
related to the negativeness of the energy density of the inward quantum flux in the
near-horizon region. This implies that the group velocity (with which real elementary
particles propagate) of the photon is slightly larger than the speed of light in the
Minkowski vacuum in the absence of external fields, i.e.
$|\partial\omega_\mathbf{k}/\partial\mathbf{k}| > c$.
The high-energy photons may, thus, potentially escape from the interior region of black holes.
This kind of strongly red-shifted photons reaching the observer at spatial infinity can
probably bring some information about the inner structure of evaporating black holes.

\end{document}